\documentclass{mem}
\usepackage{txfonts} 
\usepackage{balance}
\usepackage{graphicx}
\idline{0}{1}
\newcommand\lsim{\mathrel{\spose{\lower 3pt\hbox{$\mathchar"218$}}
     \raise 2.0pt\hbox{$\mathchar"13C$}}}
\newcommand\gsim{\mathrel{\spose{\lower 3pt\hbox{$\mathchar"218$}}
     \raise 2.0pt\hbox{$\mathchar"13E$}}}
\def\sc{Schwarzschild}
\def\ltsima{$\; \buildrel < \over \sim \;$}
\def\lsim{\lower.5ex\hbox{\ltsima}}
\def\gtsima{$\; \buildrel > \over \sim \;$}
\def\gsim{\lower.5ex\hbox{\gtsima}}

\def\apjl{ApJL}
\def\apjs{ApJSS}

\def\aap{A\&A}
\def\sch{Schwarzschild}

\begin{document}

\title{
Extra--galactic jets: a hard X--ray view 
}

   \subtitle{}

\author{
G. \,Ghisellini \inst{1} 
          }

\institute{
INAF -- Osservatorio Astronomico di Brera, Via Bianchi 46,
I-23807 Merate, Italy\\
\email{gabriele.ghisellini@inaf.it}
}

\authorrunning{Ghisellini }

\titlerunning{extra--galactic jets}

\abstract{
Extragalactic jets are the most powerful persistent sources of the universe. 
Those pointing at us are called blazars.
Their relativistically boosted emission extends from radio frequencies to TeV energies. 
They are also suspected to be the sources of energetic neutrinos and high energies cosmic rays.  
The study of their overall spectrum indicates that most of the emission of powerful 
blazars is in hard X--rays or in soft $\gamma$--rays. 
In this band we can find the most powerful jets, visible also at high redshifts. 
It is found that the jet power is linked to the accretion luminosity, 
and exceeds it, especially if they produce energetic neutrinos, that require
tre presence of ultrarelativistic protons.
% It is suggested that heavy black holes in jetted sources form earlier than in jet--less objects. 
\keywords{neutrinos -- radiation mechanisms: non-thermal -- galaxies: active -- 
BL Lacertae objects: general -- gamma-rays: galaxies }
}
\maketitle{}

\section{Introduction}
About 10\% of Active Galactic Nuclei (AGN) have relativistic jets, producing radiation
across the entire electromagnetic spectrum, that is beamed along the velocity direction.
When the jets are pointing at us, the sources are called blazars.
There two flavours of them: powerful jets do have broad emission lines with the
same properties of radio--quiet sources, and are called Flat Spectrum Radio Quasars (FSRQs), 
while less powerful sources have very weak or absent emission lines, and are called BL Lacs.
Estimates of black hole masses are available now for a large number of sources, 
allowing to estimate the accretion disk luminosity in Eddington units: $L_{\rm d}/L_{\rm Edd}$.
Therefore we now know that FSRQs have disk emitting in the standard regime 
(optically thick and geometrically thin), while most BL Lacs are have radiatively inefficient disks
(Narayan et al. 1997, Narayan et al. 2000).
In this regime, the UV emission is strongly reduced (Mahadevan et al. 1997), and the disk radiation 
cannot photo--ionize the clouds responsible for the line emission 
(Ghisellini \& Celotti 2001; Ghisellini et al. 2009).
This implies that jets can be produced both when the accretion is radiatively efficient
(and the disk is geometrically thin) and when instead the disk is radiatively inefficiently 
(and is  geometrically thick).

\section{The blazar sequence revisited}

All blazars have a spectral energy distribution (SEDs) characterized
by two broad emission humps: the first is produced by synchrotron, while the second, at 
higher energies is probably due the inverse Compton (IC) process.

\begin{figure}[h]
\vskip -0.4 cm
\resizebox{\hsize}{!}{\includegraphics{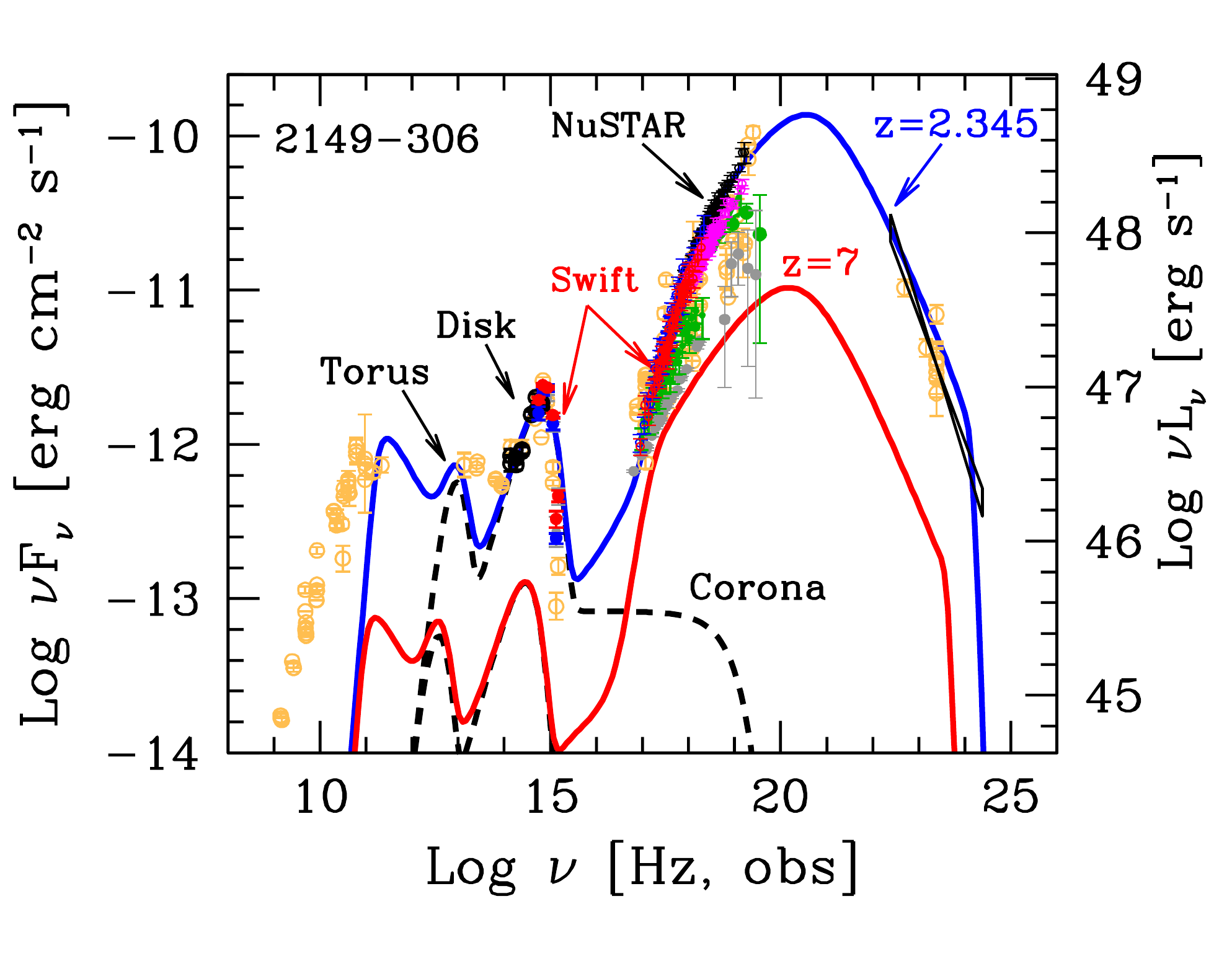}} 
\vskip -0.3 cm
\caption{
\footnotesize
The SED of the  powerful FRSQ 2149--306. 
The accretion disk emission is clearly visible, together with the IR emission 
from the molecular torus. 
The SED is dominated by the high energy bump.
}
\label{sed1}
\end{figure}

\begin{figure}[h]
\vskip -0.2 cm
\resizebox{\hsize}{!}{\includegraphics{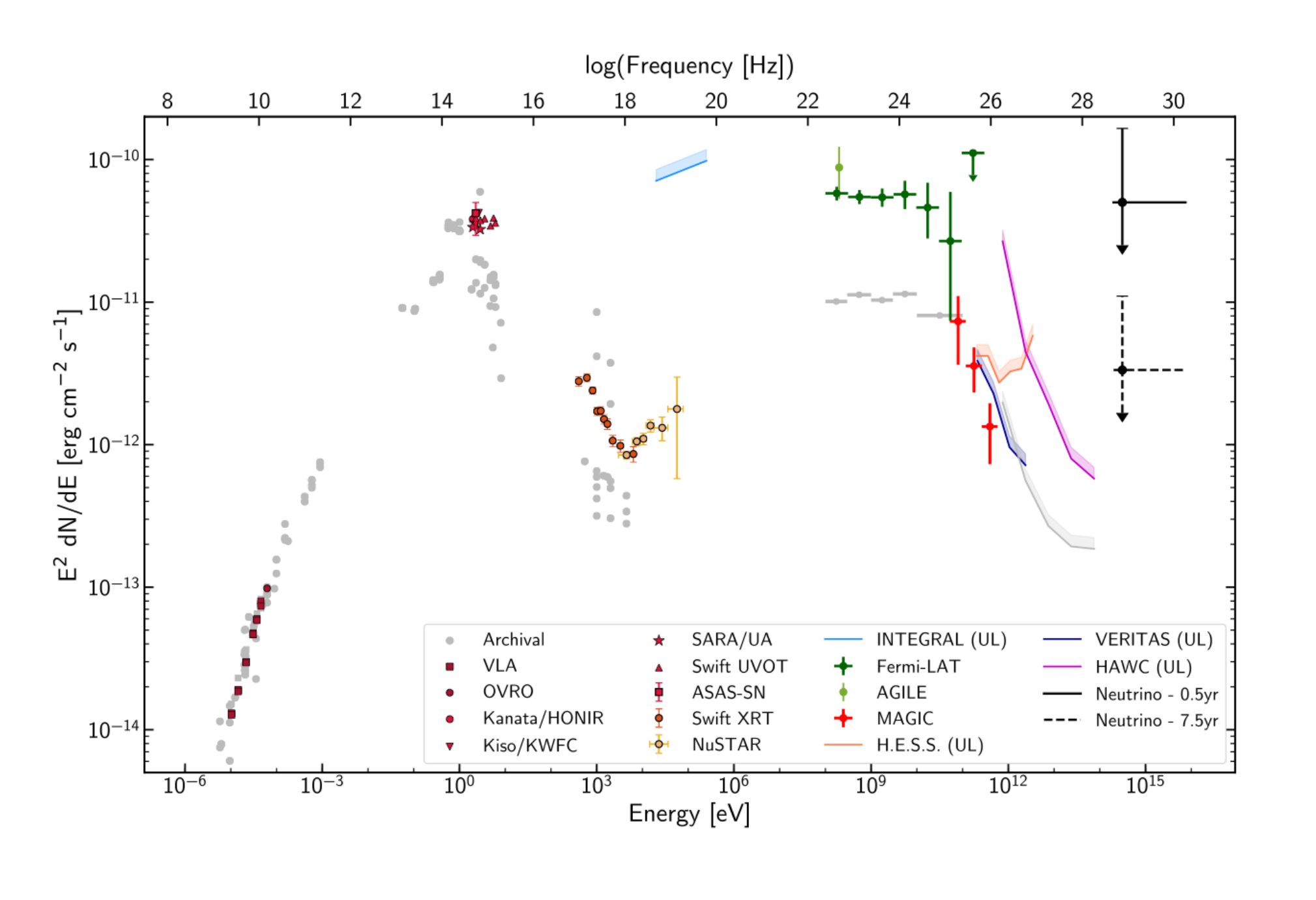}}\\
\vskip -0.5 cm
\caption{
\footnotesize
The SED of TXS 0506+365, the blazars suspected to produce high energy neutrinos.
Its SED is intermediate: the synchrotron peak is in the optical band, but the
high energy peak is at relatively high energies. 
From Aartsen et al. (2018).
}
\label{sed2}
\end{figure}

\begin{figure}[h]
\vskip -0.3 cm
\resizebox{\hsize}{!}{\includegraphics{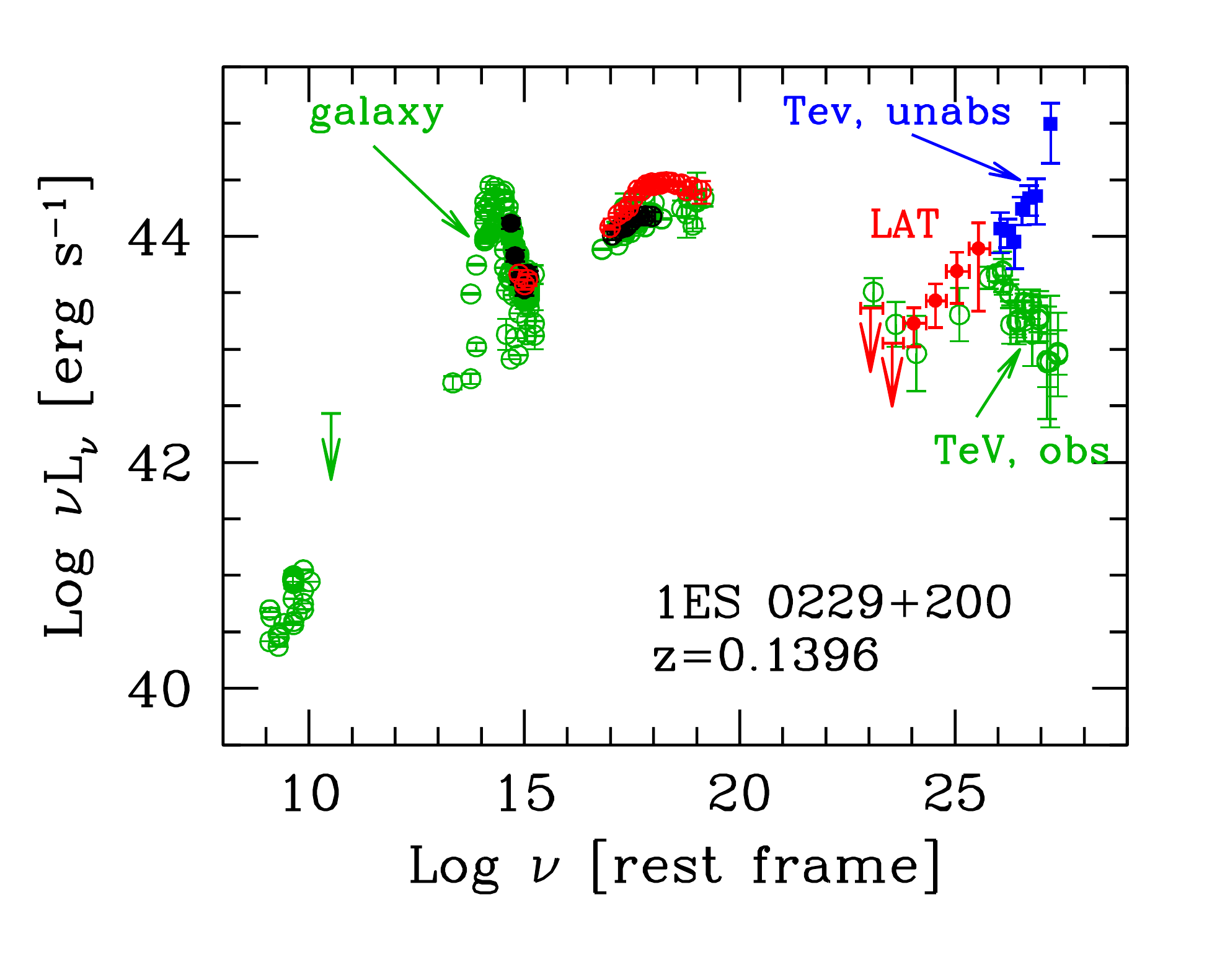}}\\
\vskip -0.7 cm
\caption{
\footnotesize
The extreme BL Lac 1ES 0229+200.  
These blazars are the best candidates to be observed in the TeV band.
Note that its TeV luminosity, once corrected for absorption, is rising in $\nu L_\nu$.
}
\label{sed3}
\end{figure}

Fig. \ref{sed1}--\ref{sed3} illustrate the SEDs of three representative 
blazars of different bolometric luminosity.
Fig. \ref{sed1} shows the SED of the powerful blazar PKS 2149--306, with $z= 2.345$.
In this case the synchrotron peak is the far IR, while the largely dominant
IC peak is around 1 MeV. 
The spectrum beyond both peaks is steep ($F_\nu \propto \nu^{-\alpha}$, with $\alpha>1$), 
and this allows the thermal emission of the accretion disk 
to be clearly visible, together with the infrared emission by the  molecular torus assumed to 
surround the accretion disk, intercepting and re--emitting a large fraction ($\gsim$50\%) of the 
accretion disk luminosity.
The curves correspond to a one--zone model and the bottom one illustrates how the SED
would appear if the redshift were $z=7$.
Even at these redshifts the source would be quite bright, especially in hard X--rays,
where the flux would be above $10^{-12}$ erg cm$^{-2}$ s$^{-1}$.

% ---------------------------------------------
% ---------------------------------------------
% ---------------------------------------------
% ---------------------------------------------
% ---------------------------------------------
% ---------------------------------------------
% ---------------------------------------------
\begin{figure}
\vskip -1.8 cm
%\resizebox{\hsize}{!}{\includegraphics{paolo.pdf}}
\resizebox{\hsize}{!}{\includegraphics{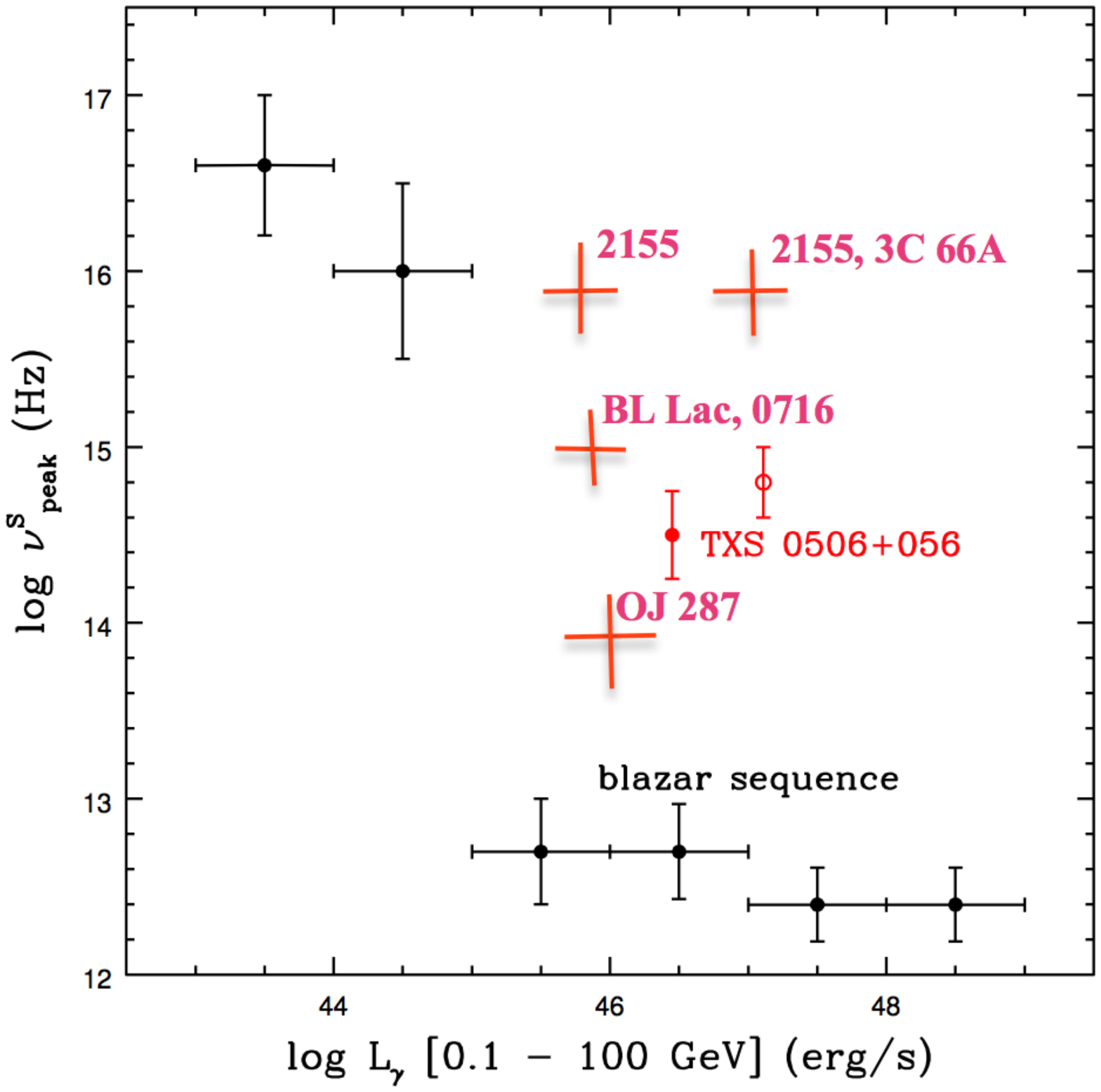}}
\vskip -1.8 cm
\caption{
\footnotesize
The peak synchrotron frequency $\nu^{\rm s}_{\rm peak}$ as a function 
of the $\gamma$--ray luminosity $L_\gamma$. 
Black points are typical values
of the blazars analyzed by Ghisellini et al. (2017).
The source TXS 0506+056 appears an outlier, but consider
that, especially during high states, other ``classical" 
BL Lacs have similar values of $\nu^{\rm s}_{\rm peak}$ and
$L_\gamma$.
Adapted from Padovani et al. (2019).
}
\label{paolo}
\end{figure}
% ---------------------------------------------

Fig. \ref{sed2} shows TXS 0506+056 (with $z=0.3365$, Paiano et al 2018), 
the source suspected to be the producer of the high energy neutrino detected by Icecube
(Aartsen et al. 2018).
Somewhat unexpectedly, this source is an intermediate blazar, with the synchrotron peak
in the optical band. 
However, it has been detected in the TeV band.
The high energy luminosity, in this kind of source,
is usually of the same order of the synchrotron luminosity.
Padovani et al. (2019) argued that this source is not a real BL Lac, but
rather a FSRQ with hidden broad emission lines.
To this aim, they show (Fig. 1 in their paper), that the synchrotron
peak frequency of this source does not agree with the blazar sequence.
However, this is not the only blazar showing, especially during flares, 
anomalous synchrotron peak frequencies with respect to their $\gamma$--ray luminosity,
as shown in Fig. \ref{paolo}. 
With PKS 2155--304, 3C 66A and OJ 287 (all classical BL Lacs),
TXS 0506+056 is in good company.

Fig. \ref{sed3} shows the SED of an extreme BL Lac.
In this context, ``extreme" means that the synchrotron peaks
in the X--ray band, and the TeV flux, once de--absorbed, is 
rising in $\nu F_\nu$.
However, the synchrotron and the high energy luminosity remain usually similar,
as in intermediate blazars.
These kind of sources are the best candidates to be detected by Cherenkov telescopes and arrays,
to test the acceleration mechanisms, to probe the extragalactic infrared and optical background,
and possibly new physics (violation of the Lorentz invariance and/or the existence of axions).

The three blazars illustrated in Figg. \ref{sed1}--\ref{sed3} 
exemplify a common trend among blazars, that we have called ``blazar sequence".
% The SEDs of blazars form a sequence:
Increasing the observed (beamed) bolometric luminosity, the frequency of both peaks 
shift to smaller values, and the high energy hump becomes more dominant.
The original blazar sequence (Fossati et al. 1998; Ghisellini et al. 1998)
was based on flux limited complete samples in radio and in X--rays, but the
$\gamma$--ray sources, detected by EGRET onboard the Compton Gamma Ray Observatory, 
were very few.

The blazar sequence was always a subject of intense debate. 
The main objection is that it reflects the outcome of selection
effects, and therefore is not a real property of blazars.
(see Giommi, Menna \& Padovani,	1999;
% The sedentary multifrequency survey - 
% I. Statistical identification and cosmological properties of high-energy peaked BL Lacs
Perlman et al. 2001;
% Surveys and the Blazar Parameter Space (about blue quasars)
Padovani et al. 2003;
% What Types of Jets Does Nature Make? A New Population of Radio Quasars (blue FSRQ con tanto X)
Caccianiga \& Marcha 2004;	
%The CLASS blazar survey: testing the blazar sequence (low LRadio and steep alpha_RX)
Ant\'on \& Browne 2005;
% The recognition of blazars and the blazar spectral sequence (Low Lradio, small v_syn)
Giommi et al. 2005;
%The sedentary survey of extreme high energy peaked BL Lacs (high Lradio high v_syn)
Nieppola, Tornikoski \& Valtaoja 2006;
% Spectral energy distributions of a large sample of BL Lacertae objects (low Lradio low v_syn)
Raiteri \& Capetti 2016;
%  Testing the blazar sequence with the least luminous BL Lacertae objects
Giommi et al. 2012; Padovani, Giommi \& Rau 2012; see also the reviews by Padovani 2007 and Ghisellini \& Tavecchio 2008).
% The discovery of high-power high synchrotron peak blazars. Poi Usato da GG+2012 per i blue quasars - Utorus vs UBLR

On the other hand, it has always been confirmed by surveys becoming deeper. 
We have recently taken advantage of the LAT detected blazars in the 3LAC 
sample (Ackermann et al. 2015)
with measured redshift (about 800 sources) to revisit the blazar sequence.
We collected their SED from the archives and divided the blazars according to their
$\gamma$--ray luminosity (Ghisellini et al. 2017).
The main results of this study was a general confirmation of the original sequence, 
with one important exception.
If we study separately FSRQs and BL Lacs, we find that FSRQ have synchrotron and high
energy peaks frequency that are approximately constant, while in BL Lacs
they shift to smaller values when increasing the observed bolometric luminosity.
On the other hand, the Compton dominance, namely the ratio of the high energy 
to the synchrotron luminosity is strongly dependent from total power, especially
in FSRQs.

% ---------------------------------------------
\begin{figure}
\vskip -0.3 cm
\resizebox{\hsize}{!}{\includegraphics{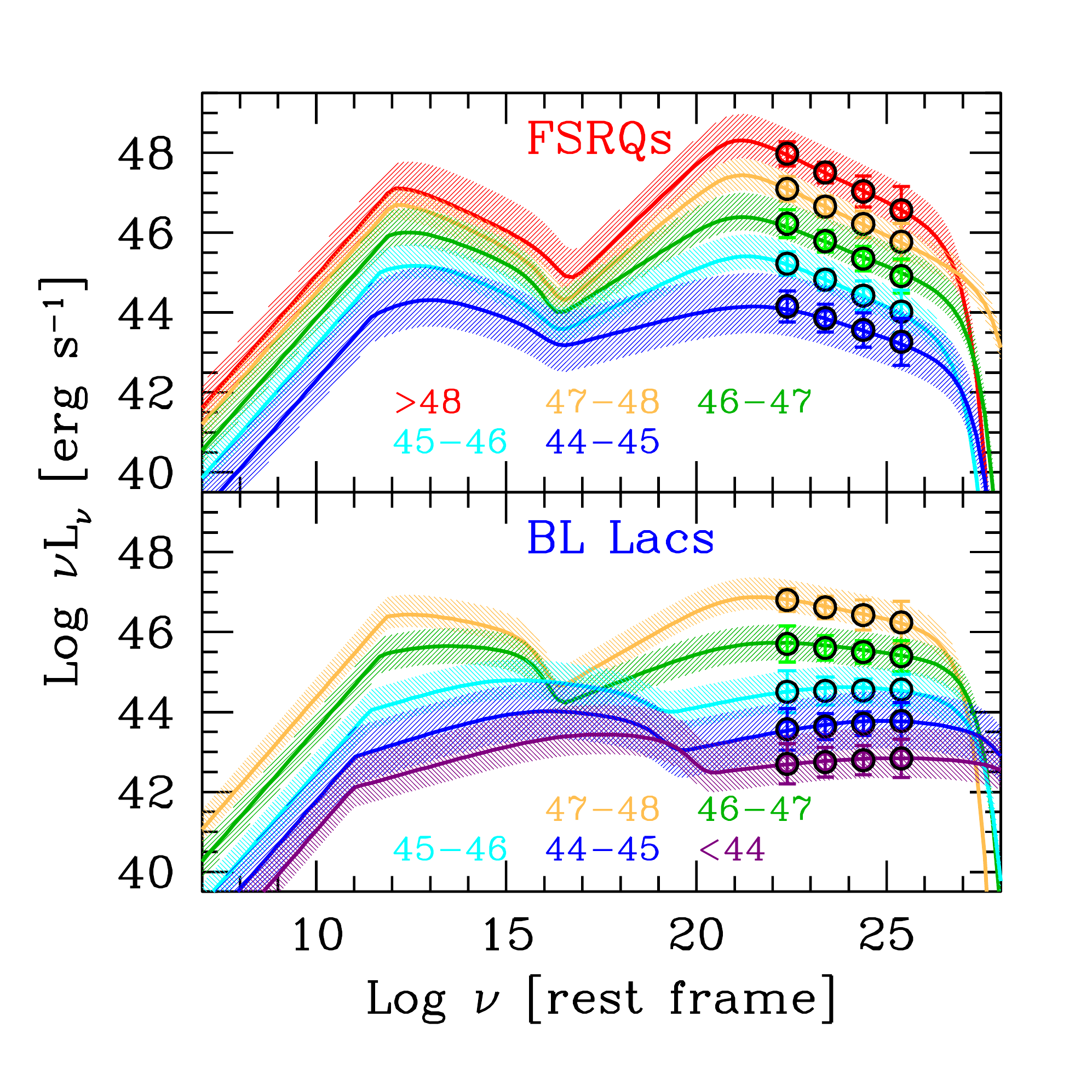}}
\vskip -0.4 cm
\caption{
\footnotesize
The new blazar sequence.
Note that for both FSRQs and BL Lacs the Compton dominance increases with the bolometric luminosity,
while only for BL Lac the emission peaks shift to smaller frequencies as the luminosity increases.
From Ghisellini et al. (2017).
}
\label{seq}
\end{figure}
% ---------------------------------------------
% ---------------------------------------------
\begin{figure}
\vskip -0.3 cm
\resizebox{\hsize}{!}{\includegraphics{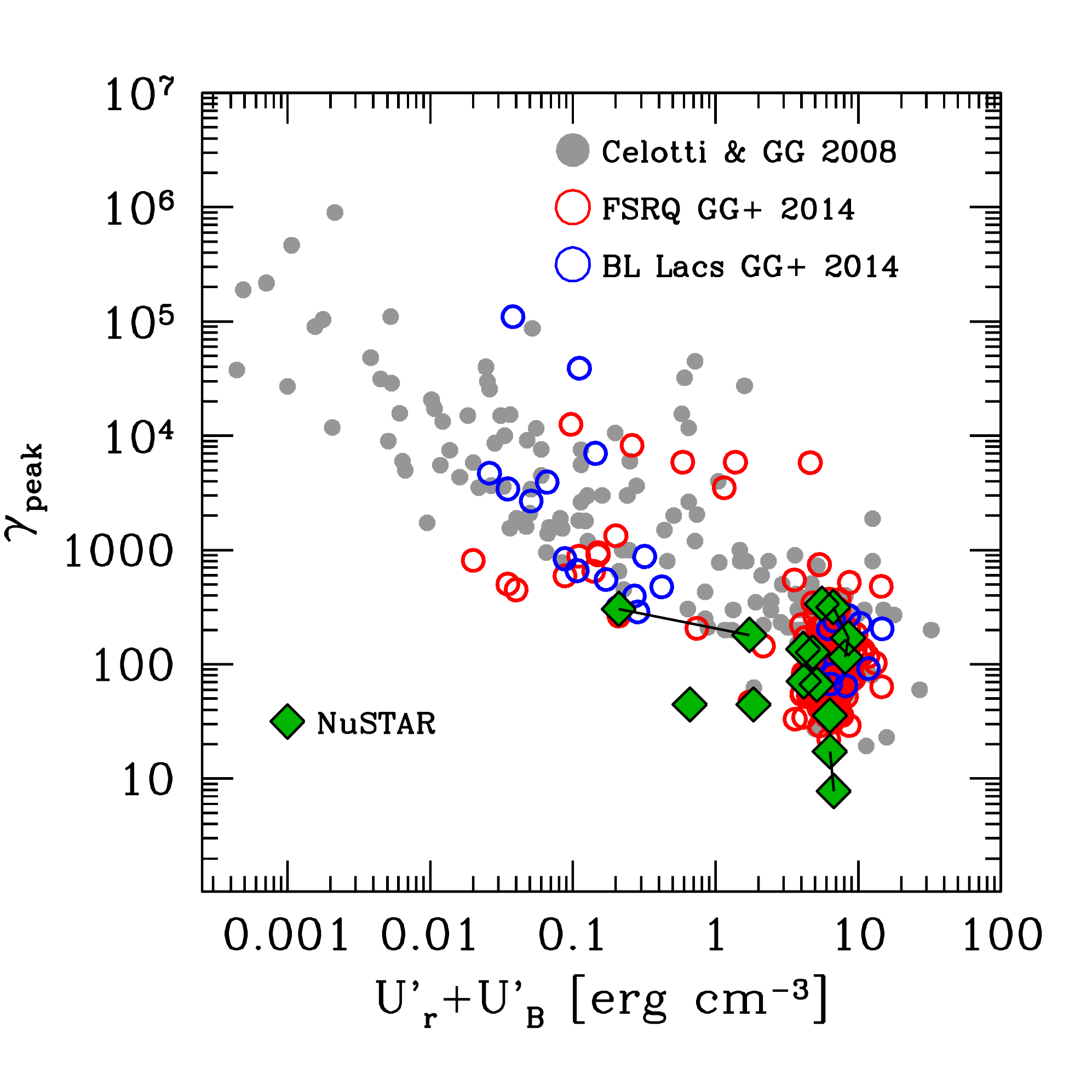}}
\vskip -0.3 cm
\caption{
\footnotesize
The random Lorentz factor $\gamma_{\rm peak}$ of the electrons emitting at both the 
synchrotron and Inverse Compton peak as a function of the comoving magnetic radiation energy density.
From Ghisellini et al. (2019).
}
\label{gpeak}
\end{figure}
% ---------------------------------------------

This agrees with the idea that the observed trend is controlled by the amount of 
radiative cooling: if in FSRQ the dissipation region is inside the broad line region (BLR)
or the molecular torus, the corresponding radiation energy density 
dominates the cooling of the emitting particles.
Both structures have a typical dimension that scales with the square root of the disk luminosity 
$R\propto L_{\rm d}^{1/2}$), making the radiation energy density $\propto L_{\rm d}/R^2$ constant.

BL Lacs, instead, have no broad lines nor obscuring tori (Chiaberge et al. 1999),
and therefore their (internally produced) radiation energy density varies.
Fig. \ref{gpeak} shows the random Lorentz fator $\gamma_{\rm peak}$ as a function
of the comoving magnetic + radiation energy density: the inverse correlation is clear.

\section{Jet power vs accretion disk luminosity}
 
Rawlings, \& Saunders (1991) were among the first to study the relation between the jet power and 
the disk luminosity, albeit in a rather indirect way. 
They found that the minimum energy contained in the radio lobe, divided by the jet lifetime,
correlated with the luminosity of the narrow lines. 
This, per se, is not surprising, since the two quantities are both dependent on redshift. 
But the important thing was that this correlation indicates that the jet power
is of the same order of the disk luminosity.
Soon later, Celotti \& Fabian (1993) devised another method to calculate the jet power, calculating
the amount of emitting electrons and the bulk Lorentz factor required to explain the 
radio emission, at the VLBI scale, from blazars.
When the first $\gamma$--ray data become available, it was realised that
i) the $\gamma$--ray luminosity dominates the electromagnetic output, and
ii) there is often a coordinated variability of the flux in different bands.
The first property called for an efficient process for producing high energy radiation,
and it was soon proposed that if the dissipating region were located within the BLR,
the comoving energy density would be boosted  by $\sim\Gamma^2$ ($\Gamma$ is the bulk Lorentz factor)
and thus would outshine the internally produced synchrotron radiation energy density.
This enhances the inverse Compton process.
The second property is the main motivation for the ``one--zone" model, that greatly simplifies
the modelling of the source, limiting the number of parameters (with respect to inhomogeneous models).

The jet power can be:

\begin{enumerate}

\item {\it magnetic:} We know that the base of the jet is optically thin.
Instead, the base of the jet of GRB is completely opaque. 
In the latter case, therefore, it is possible that the plasma is accelerated 
by its own pressure with a bulk Lorentz factor increasing linearly with the distance $R$ 
from the black hole: $\Gamma\propto R$. 
In blazars, this is not possible. 
Therefore we {\it must} invoke a magnetic acceleration.
And yet, the synchrotron luminosity is less than the inverse Compton one,
requiring that the magnetic field in the dissipating region (at $\sim 10^3$ \sch\ radii) is under equipartition.
This calls for a very fast magnetic acceleration.

\item {\it kinetic:} To find out the amount of this power component, we must 
estimate the number of particles in the jet, and thus we must assume
a radiative model.
For a leptonic, one--zone model, this is the dominant form of jet power, if we assume that there 
is one cold proton per emitting electron. 
The jet {\it cannot} be pair dominated, for two reasons:
i) without protons, there is not enough power to account for the radiation we see;
ii) a pure jet, crossing the BLR zone at high speed, would suffer a strong Compton recoil,
and would quickly decelerate.

\item {\it Radiative:} this form of jet power (let us call it $P_{\rm r}$)
is almost model--independent, and therefore is the most reliable. 
It is simply  $P_{\rm r} \sim L_{\rm obs}/\Gamma^2$ and $\Gamma$
% --value 
can be  estimated independently of the applied model, by e.g. the apparent superluminal motion.
$P_{\rm r}$ is a {\it lower limit} of the total jet power.
\end{enumerate}

% ---------------------------------------------
\begin{figure}
\vskip -0.4 cm
\resizebox{\hsize}{!}{\includegraphics{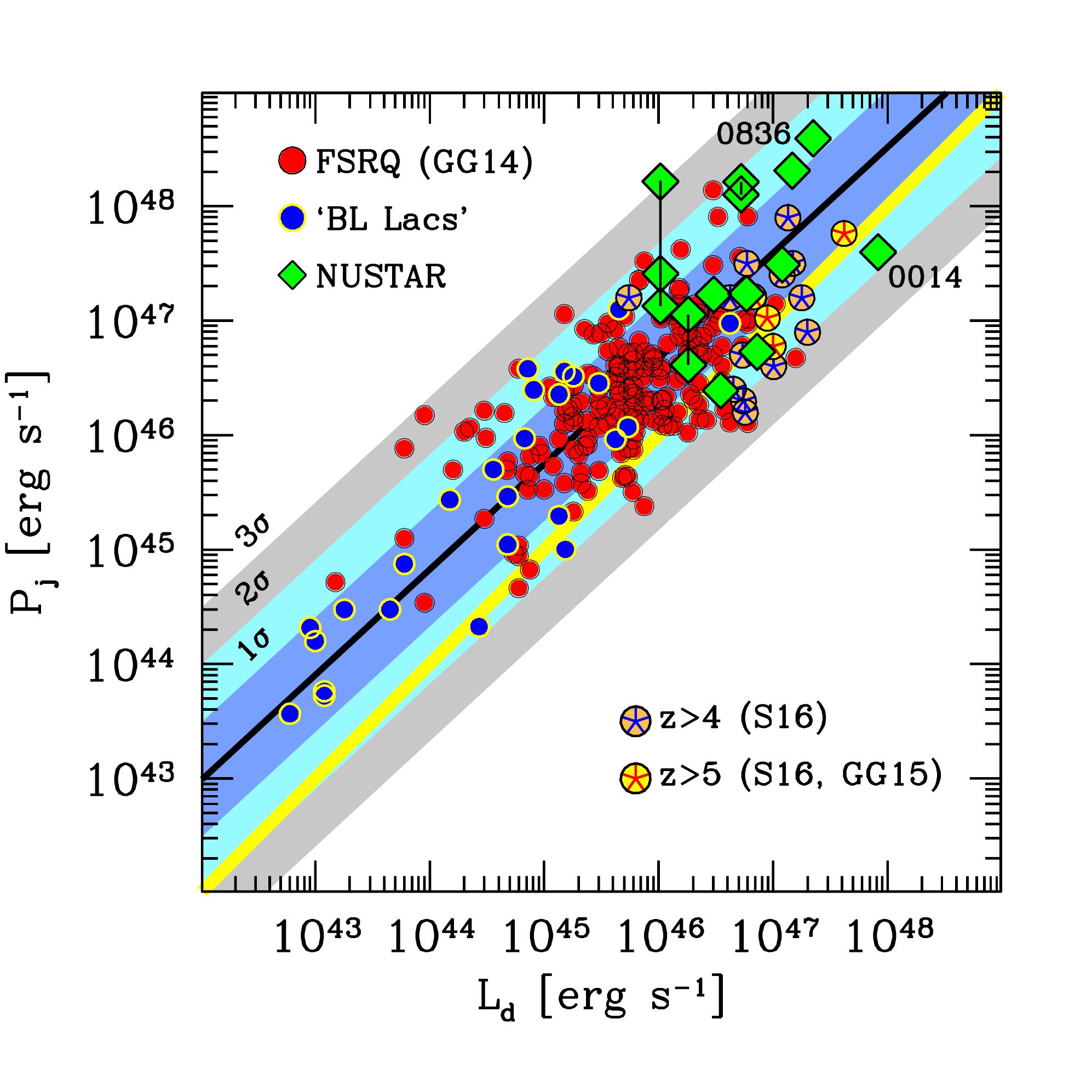}}
\vskip -0.3 cm
\caption{
\footnotesize
The jet power as a function of the accretion luminosity.
The soursed labelled ``BL Lacs" in this figure, although
consistent with the formal definition (EW$<$5\AA), 
do have broad lines in their spectrum, and should be considered as
the low luminosity tail of FSRQs.
%The sources belong to the sample studied by \cite{shaw12}, \cite{shaw13} and \cite{gg14};
%two powerful FSRQs detected by {\it NuSTAR} \cite{sbarrato16};
%a sample of $z>2$ blazars detected by {\it Swift}/BAT \cite{gg10chasing} and 
%a sample of $z>4$ blazars selected from \cite{sbarrato13}.
The yellow line is the equality line, the black line is the best fit.
From Ghisellini et al. (2014; 2019).
}
\label{pjld}
\end{figure}
% ---------------------------------------------

Fig. \ref{pjld} shows the jet power as a function of the disk luminosity for a large number of blazars.
On average, the jet power is $\sim$10 times larger than the disk luminosity, assuming
one cold proton per emitting electron.
If the latter is $L_{\rm d}=\eta_{\rm accr} \dot M c^2$, with $\eta_{\rm accr}\sim 0.1$, as it is usually
assumed, then $P_{\rm j} \sim \dot M c^2$.
This is born out also by numerical simulations, showing that, on average, $P_{\rm j} \sim 1.5\dot M c^2$
(Tchekhovskoy et al. 2012).
This strongly suggests that the jet power comes from the spin energy of the black hole
(Blandford \& Znajek 1977), and not directly from accretion
(Cavaliere \& D'Elia 2002, Ghisellini \& Celotti 2002).
Furthermore, we have now evidences that {\it jets are the result of the most powerful persistent engines 
of the universe.}

Will the $P_{\rm j} \sim \dot M c^2$ relation continue to be valid also when the accretion changes regime, from 
standard (Shakura--Sunyaev 1973 like) to radiatively inefficient 
(i.e. ADAF, Rees et al. 1982, Narayan \& Yi 1994; or ADIOS, Blandford \& Begelman 1999)?

\section{Energy crisis?}

As mentioned, there is the serious possibility that blazars are the producers of the high energy neutrinos
detected by Icecube (Aartsen et al. 2018, see also the review of Gaisser 2018).
The production of neutrinos is associated with the production of high energy $\gamma$--rays, and
it is not clear what is the type of blazar contributing the most (Resconi et al. 2017; Righi et al. 2019).   

If the association with blazars is confirmed, it implies that in their jets
there are ultra--relativistic protons with energies $\sim$20 times larger than the
observed PeV neutrino. 
Their contribution to the total power is uncertain, being a factor of a few in the case of
collision between protons and a boosted external photon field (peaking in the X--ray band, 
Tavecchio \& Ghisellini 2015),
and a few thousands if we require that hadrons are required to produce the entire SED
(Zdziarski \& B\"ottcher 2015).

If ultra-relativistic protons are indeed present in the jets of AGNs we face an
energy crisis, since their power would exceed the disk luminosity by more than
one order of magnitude.

One possible solution 
is that only a small fraction of the gravitational energy dissipated by the accreting $\dot M$ 
is transformed into heat, while the rest goes to power the jet 
(Jolley \& Kuncic 2008;  Jolley  et al. 2009). 
% \cite{jolley08}, \cite{jolley09}.
The total accretion efficiency $\eta$ is used to power the jet with an efficiency $\eta_{\rm j}$,
and to heat the disk with an efficiency $\eta_{\rm accr}$:
\begin{equation}
P_{\rm j} = \eta_{\rm j} \dot Mc^2; \,\,\, L_{\rm d}=\eta_{\rm accr} \dot Mc^2;
\,\,\, \eta=\eta_{\rm j}+\eta_{\rm accr}
\end{equation}
This does not imply that the jet power comes directly from accretion.
It may come from the black hole spin, whose rotational energy is extracted by a very
large magnetic field, sustained by a large matter density in the vicinity of $R_{\rm ISCO}$,
in turn possible because the accretion rate is much larger than what the disk luminosity
would naively suggest if we use $\eta_{\rm accr}\sim 0.1$.
% Then $P_{\rm j}= 10 L_{\rm d}$ requires $\eta_{\rm j}= 10 \eta_{\rm accr}$.
This has an important consequence: if there is a jet, the disk efficiency can be smaller 
than what is foreseen for standard accretion models.
To produce a given $L_{\rm d}$, $\dot M$ must therefore be larger, and this helps
jetted sources to have black holes that grow faster (especially at large redshifts, Ghisellini et al. 2013;
Sbarrato et al. 2015).
%[\cite{gg13}]).

% \section{Conclusions}

\begin{acknowledgements}
I thank ASI 1.05.04.95 for funding.
\end{acknowledgements}

\bibliographystyle{aa}

\begin{thebibliography}{}


\bibitem[]{} Aartsen M.G., Ackermann M., Adams J., et al. 2018, Science, 361, 1378
% Multimessenger observations of a flaring blazar coincident with high-energy neutrino IceCube-170922A

\bibitem[]{} Ackermann M., Ajello M., Atwood W.B. et al., 2015, ApJ, 810, 14 % 3LAC
           % The third catalog of Active Galactic Nuclei detected by the Fermi Large Area Telescope. 
           
\bibitem[]{} Ant\'on S. \& Browne I.W.A., 2005, MNRAS, 356, 225
   	         % The recognition of blazars and the blazar spectral sequence.

\bibitem[]{} Blandford R.D. \& Znajek R.L., 1977, MNRAS, 179, 433

\bibitem{blandford99} Blandford R.D. \& Begelman M.C., 1999, MNRAS, 303, L1  
% On the fate of gas accreting at a  low rate on to a black hole,   ADIOS         

\bibitem[]{} Caccianiga A. \& Marcha M.J.M., 2004, MNRAS, 348, 937
	         % The CLASS blazar survey: testing the blazar sequence.

\bibitem{cavaliere02} Cavaliere A. \& D'Elia V., 2002, ApJ, 571, 226
     % The Blazar Main Sequence
     
\bibitem{celotti93}  Celotti A., Fabian A.C., 1993, MNRAS, 264, 228, 
      % The Kinetic Power and Luminosity of Parsecscale Radio Jets - an Argument for Heavy Jets.

\bibitem[]{} Chiaberge M. Capetti A. \& Celotti A., 1999, A\&A, 349, 77% no torus

\bibitem[]{} Gaisser T.K., 2018, 13th Rencontres du Vietnam ``Neutrinos" 2017 proc., arXiv:1801.01551 

\bibitem[]{} Ghisellini G. \& Celotti A., 2001,  A\&A, 379, L1 

\bibitem{gg02} Ghisellini G. \& Celotti A.,  2002, 
      %Workshop: {\it Blazar Astrophysics with BeppoSAX,} % and Other Observatories, 
      % December 2001, ASI, Italy. Edited by P. Giommi, E. Massaro, G. Palumbo. 
       ESA-ESRIN, p. 257, arXiv:astro-ph/0204333  
      % Blazars, Gamma Ray Bursts and Galactic Superluminal Sources, 

\bibitem[]{} Ghisellini G. \& Tavecchio F., 2008, MNRAS, 387, 1669  % The blazar sequence: a new perspective.             

\bibitem[]{} Ghisellini G., Maraschi L. \& Tavecchio F., 2009, MNRAS, 396, L105

\bibitem{gg13} Ghisellini G., Haardt F., Della Ceca R., Volonteri M. \& Sbarrato T., 2013, MNRAS, 432, 2818 
             % The role of relativistic jets in the heaviest and most active supermassive black holes at high redshift, 
             
\bibitem{gg14} Ghisellini G., Tavecchio F., Maraschi L., Celotti A. \& Sbarrato T.,  2014, Nature, 515, 376  
            %  The power of relativistic jets is larger than the luminosity of their accretion disks
            
\bibitem[]{} Ghisellini G., Righi C., Costamante L., Tavecchio F., 2017, MNRAS

\bibitem[]{} Ghisellini G., Perri M., Costamante L., et al., 2019, subm. to A\&A

\bibitem[]{} Giommi P., Piranomonte S., Perri M. \& Padovani P., 2005, A\&A, 434, 385
             %The sedentary survey of extreme high energy peaked BL Lacs.  

\bibitem[]{} Giommi P., Menna M.T. \& Padovani P., 1999, MNRAS, 310, 465 
             % The sedentary multifrequency survey -- I. Statistical identification and  
             % cosmological properties of high--energy peaked BL Lacs.

\bibitem[]{} Giommi P., Padovani P., Polenta G., Turriziani S., DÕElia V., Piranomonte S., 2012, MNRAS, 420, 2899
             % simplified scenario
        
\bibitem[]{} Giommi P., Padovani P., 2015, MNRAS, 450, 2404 
           % A simplified view of blazars: contribution to the X--ray and $\gamma$--ray extragalactic backgrounds.

\bibitem[]{} Jolley E.J.D. \& Kuncic Z. 2008, MNRAS, 386, 989  
           % Jet-enhanced accretion growth of supermassive black holes,

\bibitem[]{} Jolley E.J.D., Kuncic Z., Bicknell G.V. \& Wagner S.,  2009, MNRAS, 400, 1521 
             % Accretion discs in blazars

\bibitem[]{} Mahadevan R., 1997, ApJ, 477, 585

\bibitem[]{} Narayan R. \& Yi I., 1994, ApJ, 428, L13  
           %  Advection-dominated accretion: A self-similar solution, 

\bibitem[]{} Narayan R., Garcia M.R., McClintock J.E., 1997, ApJ, 478, L79

\bibitem[]{} Narayan R., Igumenshchev I.V., Abramowicz M.A., 2000, ApJ, 539, 798

\bibitem[]{} Nieppola E., Tornikoski M. \& Valtaoja E., 2006, A\&A, 445, 441
             % Spectral energy distributions of a large sample of BL Lacertae objects.

\bibitem[]{} Padovani P., 2007, Ap\&SS, 309, 63 (astro--ph/0610545) 
              %  The blazar sequence: validity and predictions.

\bibitem[]{} Padovani P., Giommi P. \& Rau A., 2012, MNRAS, 422, L48 
             % The discovery of high-power high synchrotron peak blazars

\bibitem[]{} Padovani P., Oikonomou F., Petropoulou M. et al., %Giommi P., Resconi E., 
             2019, MNRAS, 484, L104

\bibitem[]{} Paiano S., Falomo R., Treves A., Scarpa R., 2018, ApJ, 854, L32
            % The Redshift of the BL Lac Object TXS 0506+056

\bibitem[]{} Perlman E.S., Padovani P., Landt H., et al.,  2001, ASPC, 227, 200 (astro--ph/0012185) 
             % Surveys and the Blazar Parameter Space.

\bibitem[]{} Raiteri C.M. \& Capetti A., 2016, A\&A, 587, A8 
             % Testing the blazar sequence with the least luminous BL Lacertae objects.

\bibitem[]{} Rawlings S. \& Saunders R., 1991, Nature 439, 138
          % Evidence for a common central-engine mechanism in all extragalactic radio sources.

\bibitem{rees82} Rees M.J., Begelman M.C., Blandford R.D. \& Phinney E.S., 1982, Nature, 295, 17  
	   % Ion-supported tori and the origin of radio jets,

\bibitem{rwsconi17} Resconi E., Coenders S., Padovani P., Giommi P. \&  Caccianiga L. 2017, MNRAS, 468, 597
	% Connecting blazars with ultrahigh-energy cosmic rays and astrophysical neutrinos

\bibitem{righi19} Righi C., Tavecchio F. \& Inoue S., 2019 MNRAS, 483, L127
	
\bibitem{sbarrato15} Sbarrato T., Ghisellini G., Tagliaferri G. et al.
           % Foschini L., Nardini M., Tavecchio F., Gehrels N., 
           2015, MNRAS, 446, 2483   % Blazar candidates beyond redshift 4 observed by Swift
	
\bibitem{ss73} Shakura N.I. \& Sunyaev R.A., 1973, A\&A, 24, 337
% Black holes in binary systems. Observational appearance, 

\bibitem{sikora94} Sikora M., Begelman M.C. \& Rees, M.J., 1994, ApJ, 421, 153  
        % Comptonization of diffuse ambient radiation  by a relativistic jet: The source of gamma rays from blazars? 

\bibitem{tavecchio15} Tavecchio F. \& Ghisellini G., 2015, MNRAS, 451, 1502
   % High-energy cosmic neutrinos from spine-sheath BL Lac jets

\bibitem{tchekhovskoy12} Tchekhovskoy A., McKinney J.C. \& Narayan R., 2012,  J. of Physics: Conf. Series, 
          372, Issue 1, id. 012040 (2012).
           % General Relativistic Modeling   of Magnetized Jets from Accreting Black Holes,

\bibitem{zdz15} Zdziarski A.A. \& Bo\"ottcher M., 2015, MNRAS, 450, L21
          % Hadronic models of blazars require a change of the accretion paradigm

\end{thebibliography}

\end{document}